\documentstyle[11pt,newpasp,twoside]{article}
\def\edcomment#1{\iffalse\marginpar{\raggedright\sl#1\/}\else\relax\fi}
\reversemarginpar

\begin{document}
\title{Using Multiwavelength Observations to Estimate the 
       Black Hole Masses and Accretion Rates in Seyfert Galaxies}
\author{James Chiang}
\affil{Joint Center for Astrophysics, 
Department of Physics,
University of Maryland, Baltimore County,
1000 Hilltop Circle, Baltimore MD 21250}
\author{Omer Blaes}
\affil{Department of Physics, University of California, Santa Barbara, CA
93106-9530}

\begin{abstract}

We model the spectral energy distribution of the type 1 Seyfert
galaxies, fitting data from simultaneous optical, UV, and X-ray
monitoring observations.  We assume a geometry consisting of a hot
central Comptonizing region surrounded by a thin accretion disk.  The
properties of the disk and the hot central region are determined by
the feedback occurring between the hot Comptonizing region and thermal
reprocessing in the disk that, along with viscous dissipation,
provides the seed photons for the Comptonization process.  The
constraints imposed upon this model by existing multiwavelength data
allow us to derive limits on the central black hole mass, the
accretion rate, and the radius of the transition region between the
thin outer disk and the geometrically thick, hot inner region in these
objects.

\end{abstract}
\vspace{-20pt}
\section{Introduction}

In a series of three papers (Chiang \& Blaes 2001, Chiang 2002, and
Chiang \& Blaes 2003), we showed that the optical/UV (OUV) and X-ray
light curves obtained from multiwavelength monitoring campaigns of
bright type 1 Seyfert galaxies could be readily explained by thermal
reprocessing of radiation from a central X-ray source by a thin
accretion disk.  These observations had been puzzling since there is
no apparent correlation between the OUV and X-ray light curves (Nandra
et al. 1998; Edelson et al. 2000) as predicted by early models of
thermal reprocessing (Krolik et al.\ 1991; Courvoisier \& Clavel 1991;
Collin-Souffrin 1991).  The resolution of these difficulties were
pointed to by Nandra et al. (2000) who noted that for the NGC 7469
observations the OUV variations {\em are} correlated with the X-ray
spectral index as well as with an extrapolation of the measured 2--10
keV fluxes up to 100 keV.  This latter result leads to the natural
conclusion that the disk flux from thermal reprocessing is driven by
the bolometric X-ray luminosity, which can differ substantially from
that seen in narrow band observations (e.g., 2--10 keV) owing to
variations in the X-ray spectral shape.

If the hard X-ray emission in type 1 Seyfert galaxies is produced by
thermal Comptonization in a hot plasma, then the X-ray spectral index
is determined by the plasma temperature and the optical depth to
Compton scattering.  Furthermore, if the thermal reprocessed emission
comprises a significant fraction of the seed photons for the
Comptonization process, then the feedback between the plasma and the
reprocessing material acts as a thermostat, regulating the temperature
of the plasma (Stern, et al. 1995). For a given optical depth, the
spectral index of the emission is then set by the strength of the
feedback, and that, in turn, is determined by the geometry of the hot
plasma and the reprocessing material and by the amount of seed photons
that are not due to reprocessing (Poutanen, Krolik, \& Ryde 1997;
Zdziarski, Lubi\'nski, \& Smith 1999).

\section{The Model}

There is an empirical relation between the Compton amplification
factor, defined as the ratio of X-ray luminosity to seed photon
luminosity entering the plasma, $A = L_x/L_s$, and the spectral index
of the thermal Comptonization continuum:
\begin{equation}
\Gamma = \Gamma_0\left(\frac{L_x}{L_s} - 1 \right)^{-1/\delta}.
\end{equation}
Here, parameter values of $\Gamma_0 = 2.06$ and $\delta = 15.6$ have
been found from fits to Monte Carlo calculations of the thermal
Comptonization process (Chiang \& Blaes 2003; see also Malzac,
Beloborodov, \& Poutanen 2001, and Beloborodov 1999). Thus, the
amplification factor $A$ is essentially determined for a given
geometry and intrinsic seed photon distribution.\footnote{There is a
slight complication in that the thermal Compton emission is not
completely isotropic, being preferentially back-scattered for the
lowest orders.}

Since the shape of the OUV spectral energy distribution (SED) of type
1 Seyfert galaxies is roughly consistent with emission from a
multitemperature thin disk, albeit with significant color corrections
(Hubeny et al., 2001), and since the X-ray emitting region must have a
fairly small size ($\la 10^{14}\mbox{cm} \sim {\cal O}(10\,GM/c^2)$)
from variability time scales, we model the geometry of the inner
regions of these objects as consisting of a hot, centrally-located
spherical plasma with a razor thin accretion disk encircling it.  Some
adjustable overlap exists between the disk and the central plasma that
allows for variation in the amplification factor, and thus, will also
allow for spectral variation.  This geometry is motivated by the disk
instability models of Shapiro, Lightman \& Eardley (1976) as well as
by more recent advection-dominated accretion flow (ADAF) models (e.g.,
Narayan \& Yi 1995).  Following these models, we will assume that the
central spherical plasma region is the disk accretion flow having made
a transition from a relatively cold, standard thin disk to a
geometrically thick, hot inner flow.

For observations of the Seyfert galaxies NGC 5548 (Magdziarz et al.\
1998), NGC 7469 (Nandra et al.\ 1998), and NGC 3516 (Edelson et
al. 2000), the combinations of ground-based optical spectra and UV
spectroscopy from IUE and/or HST provide sufficient spectral coverage
to determine shape of the OUV SED, including, crucially, the roll-over
at the higher frequencies.  In our simple picture of the accretion
disk, the emission at any given radius comprises thermal radiation
from internal viscous dissipation and from thermal reprocessing.  For
our chosen ``sphere + disk'' geometry, both of these sources of
emission obey a $\propto r^{-3/4}$ temperature distribution.
Therefore, the inner truncation radius of the thin disk, $r_{\rm tr}$,
where it makes the transition to the geometrically thick, hot central
plasma region, determines the shape of the OUV roll-over, while the
temperature at this radius sets the magnitude of the OUV SED.  Once
the disk temperature is set by the OUV observations, the source of
seed photons from both thermal reprocessing and viscous dissipation is
fully determined.

The temperature of the Comptonizing electrons, the optical depth and
the radius of the central plasma sphere can then be found by fitting
the X-ray flux and spectral index.  For a single epoch of data, this
leaves one parameter of this model unconstrained.  If one assumes a
Thomson depth of unity, then the full SED from the optical through
hard X-rays can be described with reasonable parameters (Chiang \&
Blaes 2001).  We note that this model would be completely constrained
if the high energy roll-over of the thermal Comptonization continuum
could be reliably measured. Such measurements would provide direct
estimates of both the the temperature of the Comptonizing plasma and
the overall bolometric luminosity.  In analogy with Galactic low mass
X-ray binaries, this roll-over is believed to occur at cut-off
energies of $\sim 100$ keV.  Unfortunately, there have not been any
X-ray observing missions with the effective area required to measure
the X-ray spectra at these energies on sufficiently short time scales.

Another constraint may be imposed that comes from the
expectation that the hot central plasma must be less efficient at
radiating accretion power than a standard optically thick disk would
be in its place.  This condition yields the relation
\begin{equation}
L_x < \eta \dot{M}c^2 - L_{\rm disk},
\end{equation}
where $\eta \sim 0.1$--0.4 is the expected efficiency for a thin disk
that extends all the way to the innermost stable circular orbit,
$\dot{M}$ is the accretion rate, and $L_{\rm disk} = L_{\rm
disk}(M\dot{M})$ is the luminosity of the truncated disk from internal
viscous dissipation.  Although this condition adds a constraint it
also adds yet another unknown quantity, the product $M\dot{M}$ (Chiang
\& Blaes 2003).

Multiwavelength observations of variability on time scales that are
shorter than the expected viscous times for these systems ($t_{\rm
visc} > 10^8$s) can provide additional information that can be used to
estimate the above model parameters uniquely.  On these time scales,
for typical type 1 Seyfert galaxies, we can take $M\dot{M}$ to be
constant.  Since there is one fewer constraint than there are model
parameters, we can solve for one of the model parameters as a function
of another for any given set of OUV and X-ray observations.  In figure
2 of Chiang \& Blaes (2003), we plot the plasma electron temperature,
$k_BT_e$, versus $M\dot{M}$ for each of nine epochs of the OUV/X-ray
data available for NGC 5548 (Magdziarz et al.\ 1998; Chiang \& Blaes
2003).  Assuming that a single value of $M\dot{M}$ must obtain for all
nine epochs, we can infer an upper limit based on the restriction that
the electron temperature cannot be smaller than $\simeq 30$~keV,
otherwise evidence of the thermal Comptonization cut-off would have
been seen in the epoch 7 Ginga X-ray data.  This constraint yields
$M\dot{M} \la 2.5\times 10^5\,M_\odot^2$yr$^{-1}$.  The dotted
vertical lines indicate the black hole masses $M = (1, 2, 3, 4, 5, 6,
7) \times 10^7\,M_\odot$ found for the epoch 1 data, which are the
most constraining.  Given that the black hole mass must be considered
constant as well, this yields an upper limit of $M \la 2\times
10^7\,M_\odot$.  By contrast, the mass measured for NGC 5548 using
reverberation mapping is $M \simeq 6\times 10^7\,M_\odot$ (Peterson \&
Wandel 2000).

We have applied similar analyses to data for NGC 3516 and NGC 7469
(Chiang 2002).  Although we could not estimate the disk inner
truncation radius for NGC 3516 since the OUV data do not show a high
frequency decline, we can use the observations of a broad Fe K$\alpha$
line to identify the inner truncation radius of the thin disk with the
innermost stable circular orbit. From our analyses of the
multiwavelength data, we find a black hole mass of $M \la 2 \times
10^7\,M_\odot$.  For NGC 7469, we find $M\dot{M} \sim
10^6\,M_\odot^2$yr$^{-1}$; this agrees with a value found from a
completely independent method that considers interband OUV continuum
time lags under the assumption of thermal reprocessing by a thin disk
(Collier et al. 1998).  Reverberation mapping observations place the
central black hole mass of NGC 7469 in the range $M \sim
10^6$--10$^7\,M_\odot$.  Along with our estimate of $r_{\rm tr}\sim
3\times 10^{14}$cm for the thin disk inner truncation radius, this
implies $r_{\rm tr} \ga 200 GM/c^2$ which is consistent with
observations of a narrow, unresolved Fe K$\alpha$ line as seen by ASCA
for this object (Guainazzi et al. 1994).

\end{document}